\documentclass[a4paper]{article}

\usepackage{INTERSPEECH2020}
\usepackage{multirow}

\usepackage{lipsum, soul,xcolor} 

\title{On Loss Functions and Recurrency Training for GAN-based Speech Enhancement Systems}
\name{Zhuohuang Zhang$^{1, 2}$ \thanks{Part of the work was done while Z. Zhang was an intern at Didi AI Labs, Beijing, China}, 
Chengyun Deng$^{3}$, Yi Shen$^{1}$, Donald S. Williamson$^{2}$, \\ 
\textit{Yongtao Sha}$^{3}$, \textit{Yi Zhang}$^{3}$, \textit{Hui Song}$^{3}$, \textit{Xiangang Li}$^{3}$}
\address{
$^{1}$ Department of Speech, Language and Hearing Sciences, Indiana University, USA \\
$^{2}$ Department of Computer Science, Indiana University, USA \\
$^{3}$ Didi Chuxing, Beijing, China}
\email{zhuozhan@iu.edu, $\{$shen2, williads$\}$@indiana.edu, \\
$\{$dengchengyun, shayongtao, zhangyi, songhui, lixiangang$\}$@didiglobal.com}

\begin{document}
\maketitle
\begin{abstract}
Recent work has shown that it is feasible to use generative adversarial networks (GANs) for speech enhancement, however, these approaches have not been compared to state-of-the-art (SOTA) non GAN-based approaches. Additionally, many loss functions have been proposed for GAN-based approaches, but they have not been adequately compared. In this study, we propose novel convolutional recurrent GAN (CRGAN) architectures for speech enhancement. Multiple loss functions are adopted to enable direct comparisons to other GAN-based systems. The benefits of including recurrent layers are also explored. Our results show that the proposed CRGAN model outperforms the SOTA GAN-based models using the same loss functions and it outperforms other non-GAN based systems, indicating the benefits of using a GAN for speech enhancement. Overall, the CRGAN model that combines an objective metric loss function with the mean squared error (MSE) provides the best performance over comparison approaches across many evaluation metrics.
\end{abstract}


\section{Introduction}
Speech enhancement can be used in many communication systems, e.g., as front-ends for speech recognition systems \cite{hansen1991constrained, weninger2015speech,xu2019joint} or hearing aids \cite{yang2005spectral, zhang2019objective,zhangimpact}. Many speech enhancement algorithms estimate a time-frequency (T-F) mask that is applied to a noisy speech signal for enhancement (e.g., ideal ratio mask \cite{wang2014training,gu2019neural}, complex ideal ratio mask \cite{williamson2016complex,li2020return}). Both deep neural network (DNN) and recurrent neural network (RNN) structures have been utilized to estimate T-F masks. Recent RNN approaches, such as long short-term memory (LSTM) \cite{weninger2014discriminatively} and bidirectional LSTM (BiLSTM) \cite{erdogan2015phase} networks, have demonstrated superior performance over DNN-based approaches \cite{zhang2019objective}, due to their ability to better capture the long-term temporal dependencies of speech. 

More recently, generative adversarial networks (GANs) have been investigated for speech enhancement. A number of GAN-based speech enhancement algorithms have been proposed, including end-to-end approaches that directly map a noisy speech signal to an enhanced speech signal in the time domain \cite{pascual2017segan, baby2019sergan}. Other GAN-based speech enhancement algorithms operate in the T-F domain \cite{soni2018time, fu2019metricgan} by estimating a T-F mask. Current GAN-based end-to-end systems solely use convolutional layers that have skip connections \cite{pascual2017segan, baby2019sergan}, and those implemented in the T-F domain either use only fully connected layers \cite{soni2018time} or a combination of recurrent and fully connected layers \cite{fu2019metricgan}. Convolutional and fully-connected architectures cannot leverage long-term temporal information due to the small kernal size and individual frame-level predictions, which is crucial for speech. On the other hand, recurrent-only layers do not fully explore the local correlations along the frequency axis \cite{nayem2019incorporating}. Additionally, existing GAN-based methods adopt different loss functions while using different network architectures, so it is not clear what is the best performing loss function for training such a system.

In this paper, we incorporate a convolutional recurrent network (CRN) \cite{tan2018convolutional} into a GAN-based speech enhancement system, which has not been previously done. This convolutional recurrent GAN (CRGAN) exploits the advantages of both convolutional neural networks (CNNs) and RNNs, where a CNN can utilize the local correlations in the T-F domain \cite{zhang2016deep}, and a RNN can capture long-term time-dependent information. We further extend the ``memory direction'' of the original CRN structure in \cite{tan2018convolutional} by replacing the LSTM layers with BiLSTM layers \cite{graves2005framewise}. We compare the performance of our CRGAN-based models with several recently proposed loss functions \cite{pascual2017segan, baby2019sergan, soni2018time, fu2019metricgan}, to determine the best performing loss function for GANs. The influence of an adversarial training scheme is also investigated by comparing the proposed CRGANs with a non-GAN based CRN. Furthermore, results from previous studies revealed only a small amount of improvement over some legacy approaches (i.e., Wiener filtering, DNN-based method) when they are evaluated by objective metrics \cite{pascual2017segan, baby2019sergan, soni2018time}. To better understand the benefits of GAN-based training, we additionally compare our model with recent state-of-the-art (SOTA) non GAN-based speech enhancement approaches.

The rest of the paper is organized as follows. Section \ref{sec:SE-with-GAN} provides background information on speech enhancement using GANs. In Section \ref{sec:GANframework} and \ref{sec:GAN_loss}, we describe the proposed framework and loss functions of our CRGAN model. The experimental setup is presented in Section \ref{ssec:setup} and results are provided in Section \ref{sec:results}. Finally, conclusions are drawn in Section \ref{sec:conclusions}.  

\vspace{-2mm}
\section{Speech Enhancement Using GANs}
\label{sec:SE-with-GAN}

GANs have gained much attention as an emerging deep learning architecture \cite{goodfellow2014generative}. Unlike conventional deep-learning systems, they consist of two networks, a generator ($G$) and a discriminator ($D$). This forms a minimax game scenario, where $G$ generates fake data to fool $D$, and $D$ discriminates between real 

\begin{figure}[h!]
  \centering
  \includegraphics[scale = 0.42]{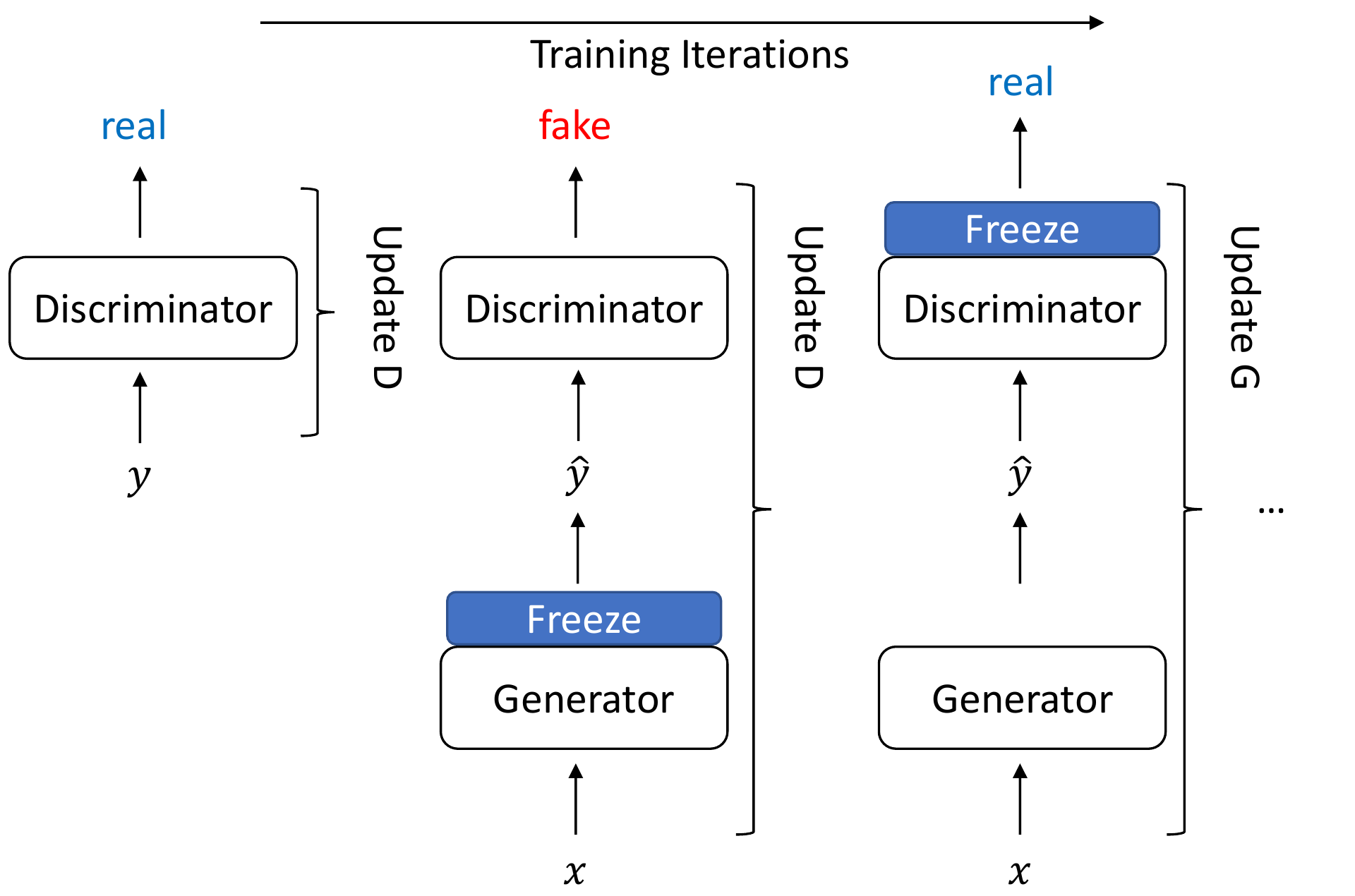}
  \caption{Training procedure for GANs. Discriminator $D$ and Generator $G$ are trained alternately. $x$ denotes the noisy log-magnitude spectrogram, $y$ represents the target (oracle) T-F mask and $\widehat{y}$ is the estimated T-F mask generated by $G$.}
  \label{fig:GAN}
  \vspace{-2mm}
\end{figure}
\noindent
and fake data. $D$'s output reflects the probability of the input being real or fake and $G$ learns to map samples $x$ from prior distribution $\mathcal{X}$ to samples $y$ from distribution $\mathcal{Y}$.

A speech enhancement system operating in the T-F domain usually takes the magnitude spectrogram of noisy speech as the input, predicts a target T-F mask, and resynthesizes an audio signal from the enhanced spectrogram. Let $\mathcal{X}$ denote the distribution of noisy log-magnitude spectrograms $x$ and let $\mathcal{Y}$ represent the distribution of target masks $y$. During adversarial training, $G$ will learn a mapping from $\mathcal{X}$ to $\mathcal{Y}$. A depiction of the GAN-based training procedure is shown in Figure. \ref{fig:GAN}. The discriminator $D$ and generator $G$ are trained alternately. In the first step of the iteration, $D$ updates its weights given the target (oracle) T-F mask $y$ that is labeled as real. Then in the second step, $D$ updates its weights again using predicted T-F mask $\widehat{y}$ generated by $G$, which is labeled as fake. Eventually in the ideal situation, given the log-magnitude noisy spectrogram as input, $G$ should be able to generate an estimated T-F mask that can fool $D$ (i.e., $D(G(x))= \text{`real'}$). Note that while $D$ is being trained, the weights in $G$ remain frozen, and vice versa.

\section{Convolutional Recurrent GAN}
\label{sec:GANframework}

\begin{figure*}[h!]
  \centering
  \includegraphics[scale = 0.6]{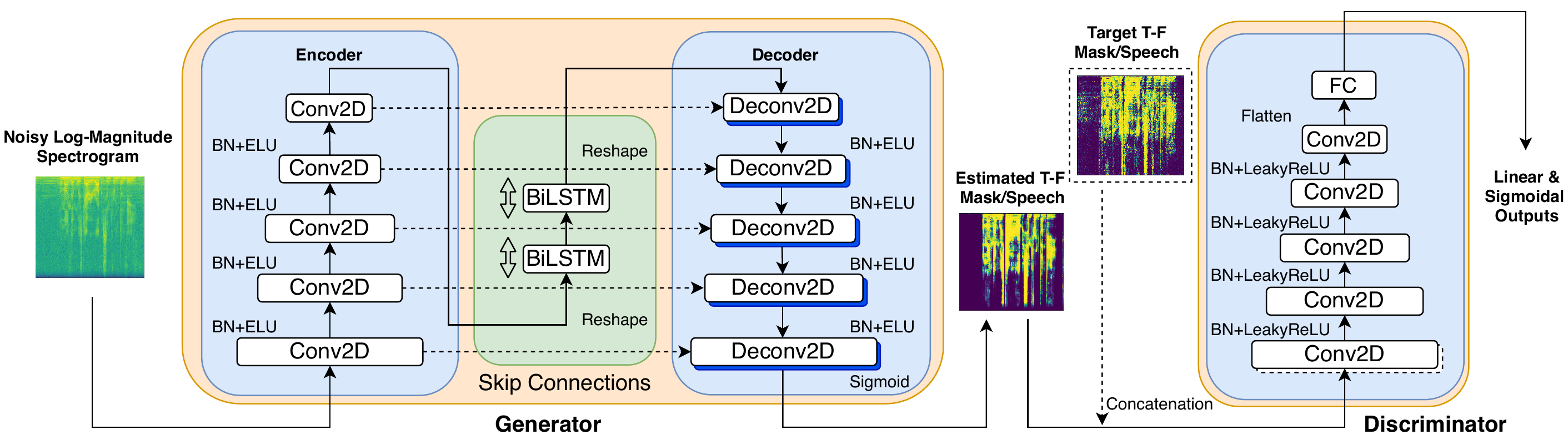}
  \caption{CRGAN structure. The generator estimates a T-F mask. The arrows between layers represent skip connections. The target T-F mask and estimated mask are provided as inputs to the discriminator for all proposed models except W-CRGAN.}
  \label{fig:CR-GAN}
  \vspace{-2mm}
\end{figure*}

The network structure for the proposed CRGAN is depicted in Figure.~\ref{fig:CR-GAN}, where $G$ has an encoder-decoder structure that takes the noisy log-magnitude spectrogram as the input and estimates a T-F mask. In the encoder, we deploy five 2-D convolutional layers to extract the local correlations of the speech signal. The encoded feature is then passed to a reshape layer, which is followed by two BiLSTM layers in the middle of the $G$ network. The recurrent layers capture long-term temporal information. The decoder is simply a reversed version of the encoder, which comprises five deconvolutional (i.e., transposed convolution) layers. We apply batch normalization (BN) \cite{ioffe2015batch} after each convolutional/deconvolutional layer.  Exponential linear units (ELU) \cite{clevert2015fast} are used as the activation function in the hidden layers, while we apply a sigmoid activation function in the output layer to estimate the T-F mask. Moreover, the $G$ network incorporates skip connections, which pass fine-grained spectrogram information to the decoder. The skip connection concatenates the output of each convolutional-encoding layer with the input of the corresponding deconvolutional-decoding layer. The network is deterministic, where the output is solely dependent on the input. 

$D$ has a similar structure as $G$'s encoder, except that we adopt leaky ReLU activation functions after the convolutional layers and there is a flattening layer after the fifth convolutional layer, which is then followed by a single-unit fully connected (FC) layer. Here $D$ gives two types of outputs, $D(y)$ or $D_l(y)$, where $D(y)$ indicates the sigmoidal output and $D_l(y)$ denotes the linear output such that $\sigma(D_l(y)) = D(y)$ with $\sigma$ being the sigmoid non-linearity. These two forms of $D$'s output are needed for the loss functions that are used to train the network.

\section{Loss Functions}
\label{sec:GAN_loss}

Existing GAN-based speech enhancement algorithms utilize different loss functions to stabilize training and improve performance. These loss functions have different benefits, but the best performing loss function is unknown. We further investigate three different loss functions within our CRGAN model to answer this question. Below, we describe three loss functions that are implemented in our model, including Wasserstein loss \cite{arjovsky2017wasserstein}, relativistic loss \cite{jolicoeur2018relativistic}, and a metric-based loss \cite{fu2019metricgan}.


\subsection{Wasserstein Loss}
The Wasserstein loss function improves the stability and robustness of a GAN model \cite{arjovsky2017wasserstein}. It is formulated as 
\begin{equation}
\label{WLoss}
\begin{aligned} 
\mathcal{L}_{D} &=  -\mathbb{E}_{y \sim \mathcal{Y}}[D_l(y)] + \mathbb{E}_{x \sim \mathcal{X}}[D_l(G(x))]\\ 
\mathcal{L}_{G} &=-\mathbb{E}_{x \sim \mathcal{X}}[D_l(G(x))],
\end{aligned}
\end{equation}
where $\mathcal{L}_{D}$ and $\mathcal{L}_{G}$ are the Wasserstein losses for the discriminator and generator, respectively. $\mathcal{L}_{D}$ maximizes the expectation of classifying the true mask as real and it minimizes the expectation of classifying a fake mask as a true one. $\mathcal{L}_{G}$ maximizes the expectation of generating a fake mask that seems real. A gradient-penalty (GP) term is included in $D$'s loss, since it prevents exploding and vanishing gradients \cite{gulrajani2017improved}:
\begin{equation}
\label{Gradient Penalty}
\mathcal{L}_{GP} = \underset{\tilde{y} \sim \tilde{\mathcal{Y}}}{\mathbb{E}}\left[\left(\left\|\nabla_{\tilde{y}} D_l(\tilde{y})\right\|_{2}-1\right)^{2}\right],
\end{equation}
where $\nabla_{\tilde{y}}D_l(\tilde{y})$ is the gradient on $D$'s linear output,  $\tilde{y} = \epsilon y + (1-\epsilon)\hat{y}$, $\epsilon$ is sampled from a uniform distribution from 0 to 1, $\hat{y}$ denotes the generated mask from $G$, and $\tilde{\mathcal{Y}}$ stands for the distribution of $\tilde{y}$.  A $L1$ loss term (i.e., $||\hat{y}_{t,f}-y_{t,f}||$) is added to $G$ to improve performance as reported in \cite{baby2019sergan}, where $\{t,f\}$ represent the time-frequency (T-F) point of the T-F mask, $y$. Thus, the discriminator loss becomes $\mathcal{L}_{D} + \lambda_{GP} * \mathcal{L}_{GP}$ and the generator loss is $\mathcal{L}_{G} + \lambda_{L1}*L1$. $\lambda_{GP}$ and $\lambda_{L1}$ serve as hyperparameters that control the weights of GP and L1 losses. We use W-CRGAN to denote this model.

\subsection{Relativistic Loss}
A relativistic loss function is adopted in \cite{baby2019sergan}, since it considers the probabilities of real data being real and fake data being real, which is an important relationship that is not considered by conventional GANs. The discriminator and generator are made relativistic by taking the difference between the output of $D$ given fake and real inputs \cite{jolicoeur2018relativistic}:
\begin{equation}
\begin{aligned} 
\mathcal{L}_{D} &=-\mathbb{E}_{(x, y) \sim(\mathcal{X}, \mathcal{Y})}[\log (\sigma(D_l(y)-D_l(G(x))))] \\ 
\mathcal{L}_{G} &=-\mathbb{E}_{(x, y) \sim(\mathcal{X}, \mathcal{Y})}[\log (\sigma(D_l(G(x))-D_l(y)))].
\end{aligned}
\end{equation}
This loss, however, has high variance as $G$ influences $D$, which makes the training process unstable \cite{baby2019sergan}. Alternatively, the relativistic average loss can be formulated as \cite{jolicoeur2018relativistic}:
\begin{equation}
\begin{aligned} 
\mathcal{L}_{D} &=-\mathbb{E}_{y \sim \mathcal{Y}}[\log (\overline{D}_{y})]-\mathbb{E}_{x \sim \mathcal{X}}[\log (1-\overline{D}_{G(x)})] \\
\mathcal{L}_{G} &=-\mathbb{E}_{x \sim \mathcal{X}}[\log (\overline{D}_{G(x)})]-\mathbb{E}_{y \sim \mathcal{Y}}[\log (1-\overline{D}_{y})],
\end{aligned}
\end{equation}
where $\overline{D}_{y}=\sigma(D_l(y)-\mathbb{E}_{x\sim\mathcal{X}}[D_l(G(x))])$ and $\overline{D}_{G(x)}=\sigma(D_l(G(x))-\mathbb{E}_{y\sim\mathcal{Y}}[D_l(y)])$. GP and L1 terms are also included to stabilize training for the discriminator and generator, respectively. R-CRGAN and Ra-CRGAN denote the models with relativistic and average relativistic loss, respectively.

\subsection{Metric Loss}
Optimizing a network using traditional loss functions may not lead to noticeable quality or intelligibility improvements. Recent approaches have thus turned to optimizing objective metrics that strongly correlate with subjective evaluations by human observers \cite{zhang2018training, kolbaek2018monaural, zhao2018perceptually}. This is adapted here, where the metric loss is defined as \cite{fu2019metricgan}:
\begin{equation}
\label{metricloss}
\begin{aligned} 
\mathcal{L}_{D} &=\mathbb{E}_{(x, s) \sim(\mathcal{X}, \mathcal{S})}[(D_l(s, s)-1)^{2}\\ & + (D_l(G(x), s)-Q^{\prime}(iSTFT(G(x)), iSTFT(s)))^{2}] \\
\mathcal{L}_{G} &=\mathbb{E}_{x \sim \mathcal{X}}[(D_l(G(x), s)-1)^{2}],
\end{aligned}
\end{equation}
where $s$ stands for the target speech spectrogram and $Q^{\prime}$ stands for the normalized evaluation metric [i.e., perceptual evaluation of speech quality (PESQ)] whose output range becomes [0,1] (1 means the best). $iSTFT$ denotes the inverse short-time Fourier transform that converts the spectrogram into a time-domain speech signal. The first input of $D$ is the enhanced signal and the second input is the clean reference signal (i.e., $D$ simulates the evaluation metric).

When using metric loss, without defining the target mask explicitly, $G$ learns a T-F mask-like representation and applies it (with an additional multiplication layer) to the noisy speech spectrogram. The generated enhanced spectrogram (i.e., $G(x)$) is then fed into $D$ to get the simulated metric scores as feedback to $G$. In such settings, $D$ learns the distribution of the actual metric score and $G$ should generate enhanced speech spectrograms with higher metric scores. We use M-CRGAN to represent our model with metric loss. Furthermore, we found that adding a mean squared error (MSE) term to $\mathcal{L}_{G}$ leads to improvements for several evaluation metrics. The generator loss becomes $\mathcal{L}_{G}+\lambda_{MSE}*||\hat{y}_{t,f}-y_{t,f}||^2$, with $\lambda_{MSE}$ being the hyperparameter that controls the MSE weight. We use M-CRGAN-MSE to represent this model.

\section{Experimental Setup}
\label{ssec:setup}

The proposed algorithm is evaluated on the speech dataset presented in \cite{valentini2016investigating}. The dataset contains 30 native English speakers from the Voice Bank corpus \cite{veaux2013voice}, from which 28 speakers (14 female speakers) are used in the training set and 2 other speakers are used in the test set. There are 10 different noises (2 artificial and 8 from the DEMAND database \cite{thiemann2013diverse}), each of which is mixed with the target speech at 4 different signal-to-noise ratios (SNRs) (0, 5, 10, and 15 dB), resulting in a total of 11572 speech-in-noise mixtures in the training set. The test set includes 5 different noises mixed with the target speech at 4 SNRs (2.5, 7.5, 12.5, and 17.5 dB), resulting in 824 mixtures. All mixtures are resampled to 16 kHz.

The log-magnitude spectrogram is used as the input feature, where we use a FFT size of 512 with 25 ms window length and 10 ms step size. For W-CRGAN and R/Ra-CRGAN, the training target is the phase-sensitive mask (PSM) defined as $M^{PSM}_{t,f} = \frac{|s_{t,f}|}{|n_{t,f}|}cos(\theta_{t,f})$ \cite{erdogan2015phase}, where $|s_{t,f}|$ and $|n_{t,f}|$ denote magnitude spectrograms of the clean and noisy speech, respectively. $\theta_{t,f}$ is the difference between the clean and noisy phase spectrograms. PSM not only estimates the magnitude but also takes phase into account. Hyperparameters are empirically set to $\lambda_{GP}=10$ and $\lambda_{L1}=200$ for W-CRGAN, R-CRGAN and Ra-CRGAN, selected based on published papers \cite{pascual2017segan,baby2019sergan}. For M-CRGAN-MSE, $\lambda_{MSE}$ is set to 4. 

The number of feature maps in the respective convolutional layers of $G$'s encoder are set to: 16, 32, 64, 128, and 256, respectively. The kernel size for each layer is (1,3) for the first convolutional layer and (2,3) for the remaining layers, with strides all set to (1,2). The BiLSTM layers consist of 2048 neurons, with 1024 neurons in each direction and a time step of 100 frames. The decoder of $G$ follows the reverse parameter setting of the encoder. For the discriminator $D$, the number of feature maps are 4, 8, 16, 32, and 64 for the respective convolutional layers. Note that the number of input channels changes when we apply the relativistic or metric loss, $D$ in this case takes in input pairs (i.e., clean and noisy pairs). All models are trained using the Adam optimizer \cite{kingma2014adam} for 60 epochs with a learning rate of 0.002 and a batch size of 60 except for M-CRGAN and M-CRGAN-MSE, where we follow a similar setup as described in \cite{fu2019metricgan} and use a batch size of 1. In M-CRGAN and M-CRGAN-MSE, the time step of the Bi-LSTM layers changes with the number of frames per utterance and each epoch contains 6000 randomly selected utterances from the training set. 

We compare our proposed system with the same network structure that does not have  recurrent layers in the generator (i.e. remove the BiLSTM) to quantify the impact that recurrent layers have on performance. These systems are denoted as W-CGAN, R-CGAN, Ra-CGAN, M-CGAN and M-CGAN-MSE. We also develop a CRN with MSE loss to investigate the influence of the GAN training scheme, and a CNN-MSE model (i.e., remove the BiLSTM) to test the influence of RNN layers on non GAN-based systems. We additionally compare with state-of-the-art non GAN-based speech enhancment systems to investigate whether it is beneficial to apply GAN training for speech enhancement. We implemented two RNN-based systems, including a LSTM-based approach and a BiLSTM-based approach. Both RNN-based speech enhancement systems have two layers of RNN cells and a third layer of fully connected neurons. In each recurrent layer, there are 256 LSTM nodes or 128 BiLSTM nodes for each direction, with time steps set to 100. The output layer has 257 nodes with sigmoid activation functions that predict a PSM training target. The MSE is used as the loss function and RMSprop \cite{tieleman2012lecture} is applied as the optimizer. The other settings are identical to the CRGAN approach. These two approaches share similar network architectures and training schemes that were mentioned in \cite{weninger2014discriminatively, erdogan2015phase}, where these approaches utilize RNN-based structures and achieve better performance than traditional DNN based approaches \cite{zhang2019objective}. 

The enhanced speech signals are evaluated using several objective metrics, including PESQ \cite{rix2001perceptual} (from -0.5 to 4.5), the short-term objective intelligibility (STOI) \cite{taal2010short} (from 0 to 1), CSIG (signal distortion), CBAK (intrusiveness of background noise) and COVL (overall effect) \cite{hu2007evaluation} (from 1 to 5).

\section{Results}
\label{sec:results}

The results for the different systems\footnote{We use author provided source code, when available, for comparison  approaches. The results for \cite{fu2019metricgan} are  different from the original results possibly due to differences with training hyperparameters (e.g. the authors in \cite{fu2019metricgan} use 400 epochs). We attach original results if code is not available.} are shown in Table \ref{results}, where all the algorithms are able to improve speech quality over unprocessed noisy speech signals. We first compare the performance of our model with state-of-the-art GAN-based systems. The results indicate that our best approach achieves much better performance in speech quality (e.g., PESQ: 2.92 in M-CRGAN-MSE) when compared to the best performing GAN-based system (e.g.~PESQ: 2.57 in RaSGAN). Similar results occur for intelligibility (e.g. STOI). This also occurs when the same loss function is used (e.g.~MetricGAN). The significant improvements in CSIG, CBAK and COVL also indicate that our proposed systems better maintain speech integrity while removing the background noise. We also include some non GAN-based systems (i.e., LSTM, BiLSTM, CRN-MSE and CNN-MSE) to verify that GAN-based systems are beneficial. It is interesting to see that all the existing GAN-based speech enhancement systems, when compared to non GAN-based systems, achieve lower scores in both enhanced speech quality and intelligibility (i.e., PESQ and STOI scores). Their enhanced speech also has a greater degree of signal distortion (CSIG) and more intrusive noise (CBAK). By comparing the performance of the proposed CRGAN models with these non GAN-based systems, the proposed models (i.e., Ra-CRGAN, M-CRGAN and M-CRGAN-MSE) tend to achieve better performance across nearly all metrics (except that Ra-CRGAN achieves lower but similar performance with BiLSTM and CRN-MSE in CSIG and COVL). This indicates that GAN-based systems can outperform non GAN-based systems, when local and temporal information are considered in conjunction with appropriate loss functions. We also notice that superior performance can be achieved by our proposed CRGAN models compared to the CRN model alone without the GAN framework, which reveals that adversarial training is beneficial for speech enhancement. 

We provide results of the proposed CRGAN-based models with different loss functions and results of these models without the recurrent layers, to quantify the importance of recurrent layers. When comparing the results, we observe a dramatic decrease in terms of speech quality (e.g., PESQ: 2.38 in Ra-CGAN and 2.81 in Ra-CRGAN), intelligibility and overall effect, which implies that recurrent layers are crucial and beneficial to GAN-based systems. We also notice that for CRN, the influence of recurrent layers is not as significant as our proposed CRGAN-based systems, suggesting that the CRGAN-based systems are more sensitive to the recurrent layers.

Among our proposed models, the CRGAN that is trained on metric loss with an additional MSE term yields the best performance across nearly all metrics (except that M-CRGAN achieve better performance in CBAK). Suggesting that a metric loss is the most beneficial loss function among the proposed ones for a CRGAN-based speech enhancement system. The CRGAN with relativisitc average loss achieves comparable performance.

\begin{table}[t!]
\caption{Results for the enhancement systems. Best scores are highlighted in \textbf{bold}. (* indicates previously reported results.)}
\vspace{-2mm}
\label{results}
\centering 
\small
\scalebox{0.86}{
\begin{tabular}{lccccc}
\hline
\multicolumn{1}{l|}{\textbf{Setting}} & \textbf{PESQ} & \textbf{STOI}  & \textbf{CSIG} & \textbf{CBAK} & \textbf{COVL} \\ \hline
\multicolumn{1}{l|}{Noisy} & 1.97 & 0.921 & 3.35 & 2.44 & 2.63 \\ \hline
\multicolumn{6}{c}{\textbf{GAN-based Systems}} \\ \hline
\multicolumn{1}{l|}{SEGAN \cite{pascual2017segan}} & 2.31 & 0.933 & 3.55 & 2.94 & 2.91 \\ \hline
\multicolumn{1}{l|}{MMSEGAN \cite{soni2018time} *} & 2.53 & 0.930 & 3.80 & 3.12 & 3.14 \\ \hline
\multicolumn{1}{l|}{RSGAN} & 2.51 & 0.937 & 3.82 & 3.16 & 3.15 \\
\multicolumn{1}{l|}{RaSGAN \cite{baby2019sergan}} & 2.57 & 0.937 & 3.83 & 3.28 & 3.20 \\ \hline
\multicolumn{1}{l|}{MetricGAN \cite{fu2019metricgan}} & 2.49 & 0.925 & 3.81 & 3.05 & 3.13 \\ \hline
\multicolumn{6}{c}{\textbf{Non GAN-based Systems}} \\ \hline
\multicolumn{1}{l|}{CNN-MSE} & 2.64 & 0.927 & 3.56 & 3.08 & 3.09 \\  \hline
\multicolumn{1}{l|}{LSTM \cite{weninger2014discriminatively}} & 2.56 & 0.914 & 3.87 & 2.87 & 3.20 \\ \hline
\multicolumn{1}{l|}{BiLSTM \cite{erdogan2015phase}} & 2.70 & 0.925 & 3.99 & 2.95 & 3.34 \\ \hline
\multicolumn{1}{l|}{CRN-MSE \cite{tan2018convolutional}} & 2.74 & 0.934 & 3.86 & 3.14 & 3.30 \\ \hline
\multicolumn{6}{c}{\textbf{Proposed CRGAN without Recurrent Layers}} \\ \hline
\multicolumn{1}{l|}{W-CGAN} & 2.29 & 0.920 & 2.60 & 2.88 & 2.42 \\ \hline
\multicolumn{1}{l|}{R-CGAN} & 2.33 & 0.916 & 2.92 & 2.81 & 2.58 \\
\multicolumn{1}{l|}{Ra-CGAN} & 2.38 & 0.917 & 2.97 & 2.87 & 2.64 \\ \hline
\multicolumn{1}{l|}{M-CGAN} & 2.59 & 0.927 & 3.68 & 3.15 & 3.11 \\
\multicolumn{1}{l|}{M-CGAN-MSE} & 2.66 & 0.926 & 3.89 & 3.05 & 3.27 \\ \hline
\multicolumn{6}{c}{\textbf{Proposed CRGAN with Different Losses}} \\ \hline
\multicolumn{1}{l|}{W-CRGAN} & 2.60 & 0.930 & 3.35 & 3.09 & 2.97 \\ \hline
\multicolumn{1}{l|}{R-CRGAN} & 2.72 & 0.932 & 3.67 & 3.09 & 3.17 \\
\multicolumn{1}{l|}{Ra-CRGAN} & 2.81 & 0.936 & 3.72 & 3.16 & 3.25 \\ \hline
\multicolumn{1}{l|}{M-CRGAN} & 2.87 & 0.938 & 4.11 & \textbf{3.32} & 3.48 \\
\multicolumn{1}{l|}{M-CRGAN-MSE} & \textbf{2.92} & \textbf{0.940} & \textbf{4.16} & 3.24 & \textbf{3.54} \\ \hline
\end{tabular}
}
\vspace{-5mm}
\end{table}

\section{Conclusions}
\label{sec:conclusions}

In this study, we propose a novel GAN-based speech enhancement algorithm with a convolutional recurrent structure that operates in the T-F domain. Results show that our proposed models outperform other state-of-the-art GAN-based and non GAN-based speech enhancement systems across an array of evaluation metrics, indicating that it is promising to use a GAN framework for speech enhancement. We conclude that the introduction of recurrent layers is important for our CRGAN model. We also investigate the influence of the GAN training scheme and different loss functions.  The metric loss greatly improves performance. By combining metric and MSE loss functions, the CRGAN approach achieves even greater performance.  

\bibliographystyle{IEEEtran}

\bibliography{mybib}

\end{document}